\documentstyle[12pt,aasms4,epsf]{article}

\def\folio{\ifnum\pageno<2\nopagenumbers\else\number\pageno\fi}
\newtoks\headline \headline={\hss\twelverm\folio\hss} 
\newtoks\footline \footline={{\hfil}} 
\font\mathbf=cmmib10 scaled 1000             

\def\ref{\par\noindent\hangindent=2pc \hangafter=1 }
\def\amin{\ifmmode^{\prime}\else$^{\prime}$\fi}
\def\asec{\ifmmode^{\prime\prime}\else$^{\prime\prime}$\fi}

\def\cappage #1 #2 #3 {\vfill\eject\pageno=#1 
\vglue 10 true in minus 10 true in \noindent{\bf Figure #2.} #3}
\def\ee #1 {\times 10^{#1}}
\def\ut #1 #2 { \, \hbox{#1}^{#2}}
\def\u #1 { \, \hbox{#1}}
\def\lsol{\, \hbox{$\hbox{L}_\odot$}}
\def\msol{\, \hbox{$\hbox{M}_\odot$}}
\def\kms {\, \hbox{km}\,\hbox{s}^{-1}}

\let\grad=\nabla
\def\cross{{\bf \times}}
\def\curl #1 {\grad \cross #1}
\def\div #1 {\grad \cdot #1}

\tighten
\singlespace

\def\msol   {\hbox{$M_\odot$}}                  
\def\kms    {\hbox{km{\hskip0.1em}s$^{-1}$}}    

\begin{document}

\title{The Galactic Center: An Interacting System of Unusual Sources}

\author{F. Yusef-Zadeh}
\affil{Department of Physics and Astronomy, Northwestern University,
Evanston, Il. 60208 (zadeh@nwu.edu)}

\author{F. Melia  }
\affil{Department of Physics and Steward Observatory, The University of Arizona
Tucson, AZ 85721; (melia@as.arizona.edu)}

\author{ M. Wardle}
\affil{Special Research Centre for Theoretical Astrophysics, University of Sydney,
Sydney, NSW 2006 Australia (wardle@physics.usyd.edu.au)}

\begin{abstract}

The region bounded by the inner tens of light years at the center of the
Milky Way contains five principal components that coexist within the
central deep gravitational potential well.  These constituents are a black
hole candidate (Sgr~A*) with a mass equivalent to $2.6\pm0.2\times 10^6$
suns, a surrounding cluster of evolved stars, a complex of young stars,
molecular and ionized gas clouds, and a powerful supernova-like remnant.
The interaction of these components is responsible for many of the
phenomena occurring in this complex and unique portion of the Galaxy.
Developing a consistent picture of the primary interactions between the
components at the Galactic Center will improve our understanding of the
nature of galactic nuclei in general, and will provide  with a better
defined set of characteristics of black holes.  For example, the accretion
of stellar winds by Sgr A* appears to produce far less radiation than
indicated by estimates based on models of galactic nuclei.

\end{abstract}

\vfill\eject

Sgr A$^*$ is a bright, compact radio source at the dynamical center of 
the Galaxy which was discovered 25 years  ago (1).
This object is
 a very strong candidate for 
a massive black hole and is the anchor about which stars and gas in 
its vicinity orbit.  Sgr A* is embedded within two clusters of massive 
and evolved stellar systems orbiting with increasing velocity 
dispersion toward it based on stellar radial velocity measurements
(2--5) 
providing a measure of the gravitational 
potential of the central mass.  Based on a remarkable set of stellar 
proper motion data acquired over six years measuring the motion of 
stars down to a field as small as 5 light days from Sgr A$^*$, a 
central dark mass concentration of
 $2.6\pm0.2\times 10^6\;M_\odot$ 
(6--9) 
has been found to lie within the inner 0.015 pc of the Galactic 
center. (0.1$''$ corresponds to 800 Astronomical Unit (A.U. is defined as the 
average distance between the Eath and the Sun)  or
1.2$\times10^{16}$cm or
about 4.6 light days or about 4$\times10^{-3}$ pc at the Galactic center distance 
assumed to be 8 kpc away.) The
inferred distribution of matter as a 
function of distance from Sgr A* (Fig. 1) (10), 
and the stellar 
velocity dispersion measurements are consistent with Keplerian motion 
(Fig. 2) (11).  The stellar kinematics within $\sim 0.01$ pc 
(3$\times10^{16}$ cm) of the Galactic center are  dominated by 
underluminous matter, probably a massive black hole, and this is 
arguably the most accurate determination of the presence of dark 
matter within the nuclei of galaxies, except perhaps for NGC 4258
(12).

However, showing that the Galactic center must contain a centralized mass
concentration does not require  that this dark matter
is in the form of a compact object with a few million solar masses
(as reference 13 had predicted).
It does not even imply that the unusual radio source Sgr A* must
be associated with it; but it is possible to demonstrate that
Sgr A* is probably not stellar.  This is based on
the fact that a heavy object in dynamical equilibrium with the
surrounding stellar cluster will move slowly, so that a failure
to detect proper motion in Sgr A* may be used to provide an independent
estimate of its mass.  In fact, such measurements have been 
carried out using the Very Large Array (VLA) of radio telescopes   for about 17 years
(14). 
More recently, similar measurements using Very Long Baseline Array (VLBA)
derived
a lower mass limit of $\sim 1000\;M_\odot$, which appears to rule
out the possibility that Sgr A* is a pulsar, a stellar binary,
or a similarly small object (15).

Still, VLBA images of Sgr A* with milliarcsecond resolution (16) show that at
a wavelength, $\lambda$, of  7 mm, its radius  is $0.76\pm0.04$ 
mas, or roughly $6.2\times 10^{13}$ cm ( about 4 A.U.), much 
smaller than the present limiting region within which the $2.6\times 
10^6\;M_\odot$ are contained.  So the dark matter may be distributed, 
perhaps in the form of white dwarfs, neutron stars, or $\sim 
10\;M_\odot$ black holes (5).  However, the 
latest stellar kinematic results appear to rule out the first two 
possible constituents.  Reference 6 argue that a 
distribution of neutron stars in equilibrium with the central 
gravitational potential should have a core radius somewhere between 
$0.15$ and $0.3$ pc,  larger than the value of $\le 0.07$ 
pc derived from the velocity data.  The same holds true for a 
population of white dwarfs.  Moreover, the neutron stars  would
presumably have been formed with a substantial ``kick'' and 
may not remain bound to the nucleus.  Thus, as long as the dark 
matter distribution is in equilibrium, the only viable alternative to 
the massive black hole paradigm may be a distributed population of 
$\sim 10\;M_\odot$ black holes.  Whether or not such a concentration 
is stable against mergers that would eventually produce a single 
massive object is still an open question, though reference (17) has 
argued that the density of dark matter in the Galactic center is so 
high ($>10^{12}\;M_{\odot}$ pc$^{-3}$) that its lifetime as a stable 
cluster could not exceed $\sim 10^8$ years, much less than the age of 
the Galaxy.

The presence of dark matter centered on Sgr A* links the Galactic 
center to the broader class of active galactic nuclei (AGNs), in which 
a massive black hole is thought to dominate the dynamics and 
energetics of the nuclear region.  A second fascinating characteristic 
of our Galaxy that it shares with
AGNs is the existence of 
fragile molecules in a ring of neutral gaseous material orbiting only a few 
parsecs from the center, not unlike the parsec-sized obscuring tori 
invoked to explain some features of AGNs (18).  The 
picture  that has emerged from a suite of multi-wavelength observations 
is that this molecular ring (also known as the Circumnuclear Disk or CND) 
with a mass of $>10^4$ \msol, is clumpy, and is rotating around a 
concentrated cluster of hot stars (IRS 16) with a velocity of about 
110 \kms (Fig. 3A) (19--21).  
Most of the far-infrared (IR) luminosity of the CND 
can be accounted for by this cluster of hot, helium emission line 
stars (22).  The IRS 16 complex consists of 
about two dozen blue stellar components at 2$\mu$m and appears to be 
the source of a strong wind with velocity of order 700 \kms and an 
inferred mass loss rate $4\times10^{-3}$ \msol$\;$ yr$^{-1}$ (23--25).	  
These blue stellar sources are themselves embedded within a cluster of 
evolved and cool stars with a radial density distribution r$^{-2}$ 
from the dynamical center of the Galaxy.  Unlike the distribution of 
the evolved cluster members, which extend over the central 500 pc of 
the Galactic bulge, the hot stars of the IRS 16 complex are 
concentrated within the inner pc of the Galaxy.

Within the cavity of molecular gas in the CND lies the ionized gas 
known as Sagittarius A West (Sgr A West) which appears as a three-arm 
spiral-like structure (with north, east, and west arms) orbiting about 
Sgr A$^*$, the IRS 16 cluster and the peak of the distribution of 
evolved stars (Fig. 3A).  The kinematics of ionized gas surrounding Sgr A$^*$ 
show systematic velocities along various components of Sgr A West 
($\approx30''$ west of Sgr A$^*$) with a radial velocity structure 
that varies regularly between $-100$ and $+100$ \kms\ in the 
south-north  direction (26--31). 
The velocity within the inner 10$''$ where there is a hole in the 
distribution of ionized gas, known as the mini-cavity, becomes 
increasingly more negative down  to $\approx-$350 \kms\ toward Sgr A$^*$ 
(32--33). These studies of gas motion over the last 20 years have
consistently indicated the presence of a large concentration of mass 
at the Galactic center.

Whereas the stars orbit randomly around the Galactic center, the 
ionized gas is part of a coherent flow with a systematic motion that 
is decoupled from the stellar orbits.  Understanding the kinematics of 
the system of ionized gas is complicated by our incomplete view of its 
three-dimensional geometry with respect to Sgr A$^*$, and is made more 
difficult by the interaction of the orbiting gas with 
non-gravitational forces due to collisions with the winds 
produced by the central cluster of hot, mass-losing stars.  Recently, 
the gas kinematics and the  geometry of the ionized flow were 
 determined by combining the transverse velocities 
measured over 9 years (Fig. 3B) and radial 
velocities of ionized gas (34). 
In the region within the central 10$''$ of Sgr A$^*$
similar proper motion results have also  been shown by
reference (35).
The predominant component of the motion in the plane of the sky is from 
east to west for many of the features, with the exception of a few where 
the velocity of ionized gas is anomalously large, possibly the result 
of the interaction between  the orbiting ionized gas and ionized stellar winds.
It appears that the 
overall flow of ionized gas in the northern arm originates in the northeast 
with negative  velocities in the orbital plane. 
 The 
ionized gas follows an orbital trajectory to the southwest as it crosses 
a disturbed region of the CND and passes behind Sgr A$^*$ before it moves to the 
northwest (21, 36) (Fig. 3C). 
The strong gravitational potential due to the large 
concentration of dark matter near Sgr A$^*$ is 
responsible for velocity gradients exceeding 
600 \kms\ pc$^{-1}$.

Moving out from the Galactic center, to a scale of 10 to 20 parsecs, 
radio continuum observations show a prominent, nonthermal continuum 
shell-like structure, known as Sgr A East.  This extended source is 
superimposed onto the thermal source Sgr A West and the CND (Fig. 4).
Low-frequency continuum observations show a decrease  in the 
brightness of the Sgr A East shell at the position of Sgr A West, 
which results from free-free absorption of the radiation from the 
former by thermal gas in the latter.  A good portion of Sgr A East 
must therefore lie behind Sgr A West (37--39).

\centerline{\bf{Sgr A$^*$: Colliding Winds and their Interaction with the Black
Hole}}

The highly compact nature of the distributed dark mass seems to suggest
that we are dealing with a pointlike object.  The gravitational field
associated with such a source intensifies rapidly with decreasing
distance to the origin, providing the necessary energy and confining
power to stress any infalling gas to temperatures in excess of a
billion degrees, which may explain the emissivity that produces the
radiation we see from Sgr A*.
 
However, suppose we play the Devil's advocate and consider the
possibility that the dark matter is not in the form of a single
massive black hole.  In that case, whatever the composition of the distributed 
dark mass concentration is, one would be left with 
the daunting task of accounting for the nature of
Sgr A* itself, without the benefit of invoking the deep gravitational
potential well of a point-like object (40).  Recently, reference (41) 
showed that a distribution of dark matter, even in the
form of $\sim 10\;M_\odot$ black holes, simply could not reproduce the 
spectrum of Sgr A*, because the gas in this region could not be squeezed
to sufficiently high densities and temperatures to produce the observed
radiative emission. However, a black hole exerts a stress on its nearby 
environment, which contains, in addition to the large scale gaseous features 
described above, rather strong winds in and around Sgr A* itself. 
The close proximity of the heavy, mass-losing stars of IRS 16 leads 
inevitably to frequent collisions between their winds, which results 
in a tessellation of broken flow segments when viewed from our 
perspective.  This process not only disrupts the spherical winds from 
these stars, but very importantly as far as the black hole is 
concerned, it facilitates the capture of gas by this object by 
reducing the plasma's kinetic energy, and thereby curtailing its 
ability to escape from the strong gravitational field.  

The portion of this wind plasma captured by the black hole falls inwards toward
the center with increasing speed as it approaches the event horizon.  In the
more energetic cores of distant galaxies, the outward push from the escaping
radiation can decelerate the flow;  not so in the Galactic center. Sgr A*
does not radiate at a level sufficient to drive this infalling gas away, and
the end result is that virtually all of the plasma within a tenth of a light
year or so is funneled into the black hole.  
Observationally, the key issue is why the infalling gas maintains a low
radiative efficiency. If we naively take the calculated accretion rate onto 
the black hole and estimate the radiative power produced by the ensuing
release of gravitational energy (42), we infer
a total luminosity in excess of what is actually measured for Sgr A*.
In fact, some estimates have it at $10^4$ or $10^5$ times the actual
observed power.  

The observations do not yet provide sufficient information
for us to identify the physics of accretion when the infalling gas
penetrates to within about $10^3$ or $10^4$ Schwarzschild radii of
the central object (no light can escape when crossed the  Schwarzschild radius of a 
black hole). The captured plasma is magnetized and highly
ionized, but it is not clear how much specific angular momentum it
carries, what the intensity of the magnetic field is, what the
relative importance of nonthermal to thermal particles is, and whether
the plasma separates into a two-temperature fluid.  As a result, a
variance of assumptions are possible (and consistent with the observations),
which results in a range of different interpretations. 
Reference (43) 
suggests that the radiation from Sgr A* 
results from shock waves  in the accreting plasma, which produce a power-law 
electron energy distribution that is truncated by strong cooling.  
This forms a ``quasi'' mono-energetic
distribution.  The overall emission, which is strictly non-thermal,
is suppressed by constraining the number density of relativistic
particles and the intensity of the magnetic field (at about 5 to 10
Gauss).
Reference (44), on the other hand, assume that
the infalling plasma eventually produces a jet of power-law electrons
whose number density varies with radius in the expulsion.  The overall
emission, which is a sum of non-thermal components, is also
suppressed by constraining the particle number density and hence
the equipartition magnetic field, both of which are assumed to be
scaled by a slowly accreting fossilized disk.
In a model developed by reference (45-46), the infalling gas
carries more angular momentum, and a disk forms with an outer radius of
about $10^5$ Schwarzschild radii.  Emission associated with the dissipation
of the additional angular momentum is suppressed if the electron
temperature is much lower than that of the protons ($T_e\ll T_p$), because  it
is the electrons that do the radiating.

One reason why the emissivity of Sgr A* is so low is that the
magnetic field in the captured plasma is  below its equipartition
value. To address this problem, reference (47) have begun
to study the behavior of such a field as the ionized gas within which
it is contained is compressed. They  find that 
whereas the rate of increase in the magnetic field intensity due to 
flux conservation depends only on the rate of compression of the gas, 
the dissipation rate is a function of the state variables and it is 
therefore not necessarily correlated with the simple equipartition
of energy. The magnetic field remains
sub-equipartition for most of the inflow, increasing rapidly only
as the gas accelerates rapidly toward the event horizon, where the
physical conditions (such as temperature and density) change more
precipitously. 

The emission in Sgr A* requires a very deep potential well, so the 
case for a massive black hole rather than a distributed dark matter 
has grown stronger.  Whether the radiation mechanism is thermal or 
non-thermal, the radiative efficiency of the infalling gas appears to 
be very low ($< 10^{-5}$).  All things considered, this low efficiency 
is probably due to either a sub-equipartition magnetic field (for 
either thermal or non-thermal models), or to the separation of the gas 
into a two-temperature plasma with $T_e \ll T_p$.  

It is important to note, however, that only a fraction of the gas available from stellar
winds of IRS 16 is actually captured by the black hole and eventually accretes towards
it.  Much of the plasma is gravitationally focused as it passes by the central potential
well, but remains unbound and continues to flow beyond the interaction zone, possibly
impacting other gaseous structures in that region.  Some evidence for this is provided
by the presence of a chain of plasma blobs (Fig. 5A) that appear to be in transit from
Sgr A$^*$ toward the ionized bar (Fig. 5B)  located south-west of the dynamical center
of the Galaxy (48-50).  A small cavity (Fig. 5)
has been carved out of the
ionized bar (51, 52), possibly due
to the impact of the collimated flow from the direction of the black hole.  Indeed,
hydrodynamical simulations suggest that the Bondi-Hoyle process responsible for the
accretion of $\sim 10^{22}$ g s$^{-1}$ by Sgr A* also produces a downstream, focused
flow with a radius very similar to that of the cavity and a mechanical luminosity
sufficient to power its radiative emission (51).

\centerline{\bf{The Cometary Tail of IRS 7: A Supergiant Star Bathed}}
\centerline{\bf{by the Radiation and Winds from IRS 16}}

The infrared source IRS 7 is a class M2I red 
supergiant lying within the projected distance of one light year from the Galactic 
center (53-54).  This source, which is a member of the evolved cluster of stars, 
has a mass-losing envelope that is being ionized externally by the 
bath of UV radiation filling the central cavity (3, 32).
Radio continuum observations have revealed a ``cometary tail'' of 
ionized gas from IRS~7 projected away from the dynamical center of the 
Galaxy (55).  The kinematics of Ne$^+$ 
emission from IRS 7's ``tail'' show conclusively that this ionized 
feature is physically associated with IRS 7 (56). 
 Recently, Keck observations show
near-IR emission 
from the tail of IRS 7 having a dust temperature of $<$200K (50). 
The detected free-free emission
from the outer 
envelope of IRS 7 (Fig. 6) is consistent with photoionization by a UV radiation 
field of the strength inferred to lie within inner 2 pcs of the galaxy 
(32, 55, 3, 56, 58), 
which imply a centrally concentrated source producing ionizing photons 
at a rate of $2\ee 50 \ut s -1 $ (59-61).

These radio and mid-IR observations suggest that the observed 
phenomena may be explained by the ionization and subsequent removal of 
the mass-losing envelope of IRS 7 by the ram pressure associated with 
the nuclear wind.  In this interacting picture between the evolved and 
young clusters, the ram pressure due to a point-like Galactic center 
wind is responsible for ablating the circumstellar envelope of IRS~7.  
The tail, however, is longer and thinner than expected (Fig. 6) if the 
expanding stellar envelope is uniform.  If this were actually the 
case, the radius  of the head of IRS 7 would be determined by the 
stand-off distance at which the ram pressures of the expanding 
envelope and the IRS 16 wind are comparable.  The opening angle of the 
tail would be larger than observed, because the effective pressure of the wind acts
only in 
the wind direction and cannot effectively confine the envelope 
perpendicular to the tail.  Instead, the expanding envelope of IRS 7 is 
probably inhomogeneous (62-63), as appears to be the 
case in other circumstellar envelopes.  The stellar envelope then 
consists of a collection of clumps of dense gas moving radially 
outwards from the central star.  The inertia of the clumps is 
sufficient that their motion is unaffected by the drag from the nuclear 
wind, which slowly ablates material from their surfaces.  The size of 
the head of the comet-like structure is then determined by the 
expansion speed of the envelope and the time scale for an individual 
clump to be ablated away.  The thin tail is produced by pollution 
(mass loading) of the IRS 16 wind by the material stripped from 
the clump surfaces (63).

\centerline{\bf{The High-Energy Emission of Sgr A East: Irradiation of Relativisitc}} 
\centerline{\bf{Particles by the Luminous Central Sources}}

The CND is a powerful source  of mid 
to far-IR continuum emission (about $2\ee 6 \lsol$), which is interpreted to be 
re-radiation by warm dust in the ring that has absorbed the same power 
in the UV (64-65).  
This implies a total UV luminosity in the region of about  $2\ee 7 
\lsol $, consistent with the radio continuum emission from Sgr A West 
and with the far-infrared line emission from the ring (66).  
On a larger scale, there is a diffuse halo of   nonthermal continuum emission 
with a diameter of about $7^{\prime}-10^{\prime}$ surrounding the oval-shaped nonthermal  
structure Sgr A East which itself lies close to CND (Figs.  3 and 4). 
The power-law energy distribution of the relativistic
electrons within the shell and the halo has a spectral index $\alpha$ 
estimated to
be 3 and 2.4, respectively, in these sources (38). 
The optical depth (defined to be the size of the emitting region divided
by the photon's mean free path length) toward Sgr A East  and the halo at low frequencies   
lead these authors  to consider a mixture of both thermal and
nonthermal gas in the halo, though displaced to the front side of Sgr A East.
The schematic diagram in Figure 7 assumes a geometry in which
the Sgr A East shell lies close to, but behind, the Galactic center
whereas the diffuse Sgr A East halo surrounds the Galactic   
center and the shell.

The strong coupling between the relativistic decay particles in Sgr 
A East, its halo  and the external UV and IR radiation from the central $1-2$ 
pc, was the motivation behind  explaining the nature of
the  EGRET $\gamma$-ray source 3EGJ1746-2852 which is 
positioned at Galactic longitude $l= 0.11^o$, and Galactic latitude $b=-0.04^0$
(67-69). 
The inverse Compton (x-ray and $\gamma$-ray)  
emissivity produced by this relativistic $e^+e^-$ population bathed by 
the IR and UV radiation is emitted isotropically from within a volume 
$V\sim 250$ pc$^3$ in Sgr A East, corresponding to a shell with radius 
$R\approx 5$ pc and thickness $\Delta R\approx 1$ pc.  The remaining 
continuum component of importance is bremsstrahlung  
resulting from the interaction between the relativistic leptons and 
the ambient nuclei.  In this model, the primary 
physical interaction accounting for the broadband spectrum of Sgr A East
(Fig. 8) is the shock wave acceleration of 
protons to relativistic energies (61) due to 
the collision of Sgr A East and its nearby molecular clouds (see below). 
 As these highly energetic particles escape from the shock wave regions, 
they  scatter with the ambient (low-energy) protons, 
which produces a proliferation of neutral and charged pions.  
The neutral pions decay and form a $\gamma$-ray spectrum, whereas the 
charged pions decay into muons and thence into electrons and 
positrons. In this model, the overall spectrum from 
Sgr A East is a superposition of the $\gamma$-rays from $\pi^0$ 
decays, synchrotron radiation by the relativistic leptons produced 
during the decay of the charged pions, bremsstrahlung emission by 
these electrons and positrons, and their Comptonization (the
process by which energetic particles boost the energy of the ambient
radiation via scattering events) of the IR and UV radiation from the central $1-2$ pc.

\centerline{\bf{Interaction of Sgr A East and the 50 \kms\ Molecular cloud}}

The relationship between the dense gas clouds in the inner 10 to 20 pc 
of the Galaxy and the CND, Sgr A$^{\mathrm{*}}$, and 
Sgr A East is central to our understanding of the current 
environment and recent history of the Galactic center.  In the last 
few years it has become apparent that Sgr A East, although lying 
almost entirely behind the center, is close enough to partially 
envelope Sgr A West and the high pressure associated with this remnant 
is sufficient to disturb the nearby 50 \kms\ cloud (70) 
and quite possibly 
the CND as described below.

The initial evidence for this interaction with the 50 \kms\ cloud came from
observations of the cloud in CO and millimeter wavelength emission from cold dust
which showed how the atomic and molecular gas curved around the nonthermal shell as
if in response to an interaction (70-74).
Further molecular line 
observations demonstrated a velocity gradient consistent with 
acceleration by Sgr A East  (75).  More recent 
studies at higher resolution with the ammonia molecule
(76) have confirmed these ideas and suggest that the gas is warmer as a 
result of this disturbance.  Supporting evidence has also been 
provided by the detection of OH masers at 1720 MHz associated with Sgr 
A East (36, 77). 
Empirically, these masers occur in the 
Galaxy where supernova remnants impact adjacent molecular clouds (78).  
There are  
theoretical reasons for this: when the density and temperature in the 
gas fall within a restricted range ($n\sim 10^5 \ut cm -3 $, $T\sim 
50$--$125 \u K $), collisions of H$_2$ molecules will invert the 1720 
MHz transition of the OH molecule (79-80).  
The production of a significant abundance of OH in the post-shock 
wave gas 
requires dissociation of water.  This can be achieved by UV
irradiation of the molecular gas, but if this is too intense, the
resultant grain heating generates a far-IR continuum that inverts the 
1665 and 1667 MHz transitions instead (80). On the other hand, 
dissociation can occur because of the irradiation of the
molecular cloud by x-rays produced by the hot gas in the interior of
the adjacent SNR (Fig. 9). These conditions, and an adequate abundance of OH, can 
be achieved in shock waves in molecular clouds (81); and in 
Sgr A East the conditions  agree with this model (72). 
Emission in IR  
rotational-vibration transitions from hot molecular hydrogen has  
been observed towards the masers in Sgr A East (77) (Fig. 10), 
although further observations are required to confirm that the 
emission comes from gas that has been heated by a shock wave rather 
than (for example) by irradiation from nearby hot stars.

\centerline{\bf{Interaction of Sgr A East with the CND}}

The suggestion that the front edge of the expanding shell may also 
have overrun the CND (39) is more indirect 
than the case of an interaction of Sgr A East with the 50 \kms\ molecular cloud,
as it is harder to detect the associated velocity 
gradients because of the clumpy and disordered nature of the CND. The 
evidence suggests that the CND is disturbed from outside, probably by 
Sgr A East.  Firstly, although the ionized gas 
associated with Sgr A West is absorbing most of the nonthermal 
emission from Sgr A East and must therefore lie in front, there is 
still nonthermal emission present at $\lambda$=90 cm toward thermally ionized gas; 
Sgr A West is therefore 
embedded within Sgr A East, but lies towards the frontmost edge.  
Second, near-IR observations have revealed a linear filament of H$_2$ 
emission located at the western edge of the CND running parallel to 
the nonthermal shell of Sgr A East.  The sheet-like morphology
association with OH maser source C at 1720 MHz (Fig. 10), and the lack of evidence for
UV 
heating in the form of thermal radio continuum or Br $\gamma$ emission, 
imply that this filament is shock-heated. 
The high-velocity of the sheet-like H$_2$ gas and the OH maser C are consistent with a 
model in which they outline the outer envelope of the CND, probably 
shock-excited externally by the nonthermal source Sgr A East (Fig. 10). 
Almost all  members of the class of OH masers observed at 1720 MHz are associated 
with nonthermal radio continuum sources. It is possible that the 
shocked gas is associated with the 50 \kms\ cloud and happens to be 
aligned fortuitously along the outer edge of the CND, but the 
morphology in several tracers appears to link it to the CND 
(39).  Finally, highly negative radial velocity H$_2$CO, OH, 
HI and HCO$^+$ absorption features, with velocities of about --190 
\kms\  (82-86) have been observed 
towards Sgr A West.  The kinematic and spatial distribution of this 
gas place it at the Galactic center (but see reference 87 for 
an alternative interpretation).  If this gas is associated with Sgr A 
West, the only plausible explanation for its highly negative velocity is that it 
has been accelerated by Sgr A East.

\centerline{\bf{Sgr A East: Is this Remnant the Result of an Interaction}}
\centerline{\bf{between a Star and the Black Hole?}}

The ample evidence that the nonthermal shell of Sgr A East is 
physically interacting with the 50 \kms\ molecular cloud 
has suggested a model in 
which an explosion occurred inside the molecular cloud and created the 
Sgr A East shell (37, 71, 74). In this scenario, the mass (6$\times10^4$ \msol)
of neutral gas that curves around 
the shell of Sgr A East has been swept up by the explosion (74).
Although resembling a SNR, Sgr A East's inferred energetics ($\sim 
4\times 10^{52}$ ergs); (74) appear to be extreme and 
have generated some uncertainty regarding this interpretation.  The 
explosion that produced Sgr A East may instead have been the tidal 
disruption of a main sequence star whose trajectory took it within ten 
Schwarzschild radii of the central object (88).  
In this picture, the gravitational field of the black hole squeezes 
the star into a long thin spike during its inward trajectory, and the 
work done by gravity is dissipated quickly into internal energy of the 
unfortunate intruder.  The energy stored in this fashion can exceed 
the binding energy of the star by several orders of magnitude, and so 
when it recedes from its location of closest approach to the black 
hole, the star expands explosively, very much like a supernova shell, 
except with a much greater energy.
Alternatively, the energetic requirements are significantly reduced if the
explosion 
that created Sgr A East expanded into a pre-existing cavity rather than 
a dense cloud.  The energy required is about  the thermal 
energy of the $T\sim 10 \u keV $, $n_e \sim 6 \ut cm -3 $ gas detected 
in x-rays (Fig.  9) by the Advanced Satellite for Cosmology and Astrophysics
(ASCA) (89),
about $5\times 
10^{51}\u erg $ -- equivalent to several normal supernovae.

The formation of  such a cavity  in the original 50 \kms\ cloud can be
described as follows.   The 
distribution of molecular emission in CO, H$_2$CO, CS, HCO$^+$ all 
indicate a lack of molecular gas at velocities around 50-60 \kms\ 
toward Sgr A West (90-92).  This suggests a scenario in which both the cavity and
the CND 
are a consequence of dynamical interaction of the original 50 \kms\ 
cloud and the gravitational potential of the inner few parsecs of the 
Galaxy.  In this model, star formation has been taking place in the 
initial 50 \kms\ molecular cloud as it sweeps through the Galactic 
center, engulfing Sgr A*.  The inner regions of the cloud are 
deflected effecting a collision between gaseous material  that pass on 
either side of Sgr A* with opposite angular momenta.  The resulting 
dissipation permits this gas to become bound to the Galactic center.  
The subsequent circularization and settling happens rapidly (93).  
This model  is consistent with the asymmetry and 
disorder of the CND, which indicate that it is perhaps a few tens of 
orbits old (94).  Meanwhile, the outer part of the 
50 \kms\ molecular cloud continues passing through the Galactic center 
on its way to its present position between 5 and 30 pc behind Sgr A 
West.  The interaction would have occurred $\sim 3 \times10^5$ years 
ago, and during this time, the progenitor of Sgr A East could have 
exploded inside the cavity that has been cleared out by the transit 
through the Galactic center.

\centerline{\bf{Summary}}

This review has examined the common thread of interaction among five sources in order
to explain the various observed phenomena in the  rich, complex and unique portion of
the center of the Galaxy.  Any self-consistent picture of this region must include an
accounting of how the five principal members of this closely packed system interact
physically with each other.  The constituents that were discussed 
are the black hole, Sgr A* (a name
derived from its identification at radio wavelengths), the surrounding cluster of
bright stars, IRS 16, the red supergiant star, IRS 7, with its attached cometary-like
tail, the Circumnuclear Disk (CND) of molecular gas, and a powerful supernova-like
remnant known as Sgr A East, which envelops many of the other objects.
In spite of the fact that the Galactic
center is totally obscured at optical wavelengths,  
the precision with which the mass of the black hole is determined 
is creating new puzzles in understanding the type of activity 
found in nuclei of galaxies that are not obscured at  optical wavelengths. 
In particular, why this
massive black hole  has such an unexpectedly low radiative efficiency.
Understanding the answer to
this puzzle will greatly improve our overall view of how the central engines (also
thought to be massive black holes) in Active Galactic Nuclei derive their power and
characteristics. 

Acknowledgments: We like to thank A. Eckart, A. Ghez. D. Roberts 
and S. Stolovy for 
their help in providing figures used here. This research was partially  supported
by NASA. 

\vfill\eject

\vfill\eject
\begin{figure}
\caption{A plot of the distribution of enclosed mass versus distance from
Sgr A$^*$.  The three curves represent the mass model for a nearly
isothermal stellar cluster with a core radius of 0.38 pc (the thick dashed
line), the sum of this cluster plus a point mass of 2.61$\pm$0.35 $\times
10^6 \msol $ (the thin continuous curve), the same cluster and a dark
cluster with a central density of 2.2$\times10^{12} \msol$ pc$^{-3}$ and a
core radius of 0.0065 pc (the thin dotted curve).  (10).  Data obtained
with the New Technology Telescope (NTT)}
\end{figure}

\begin{figure}
\caption{The projected stellar velocity dispersion versus the distance from
Sgr A$^*$.  The solid curve represents Keplerian motion due to a mass
concentrated within 0.01 pc.  (11).  Data obtained with the Keck telescope
(NTT).}
\end{figure}

\begin{figure}
\caption{(3A) A radio continuum image of ionized gas (Sgr A West) at
$\lambda$=1.2cm with its three-arm appearance, shown in blue,
superimposed on the distribution of HCN emission displayed in red
(20).  Most of the ionized gas is distributed in the molecular cavity.
(3B) The direction of ionized flow superimposed on a radio continuum
image at $\lambda$=2cm based on proper motion measurements carried out
over 9 years (34).  This pseudo-color continuum image is from the 1990
epoch data with a resolution of 0.1$''\times0.2''$ based on VLA
observations.  The boxes are regions where proper motion measurements
have been averaged and the velocities projected on the plane of the
sky are represented by red arrows.  The bright white source close to
the center of the image coincides with the position of Sgr A$^*$.
(3C) A cartoon of the ionized gas and its relationship with respect to
the CND orbiting Sgr A$^*$ represented by a black spot.  The colors
show red and blue-shifted radial velocities of molecular gas
corresponding, respectively, to negative and positive radial
velocities of the CND. The black drawings trace ionized gas at the
inner edge of the western CND as well as the edge of an intruding
cloud which apprears to be disturbing the kinematics and the
morphology of the CND.}
\end{figure}

\begin{figure}
\caption{VLA radio continuum image of the Galactic center showing
the shell-like structure of nonthermal Sgr A East (light blue and green) and the
spiral-shaped structure of thermal Sgr A West (red) at $\lambda$=6cm.  A
cluster of HII regions associated with Sgr A East are also evident to
the east of the shell.  The weak extended features (dark blue)
surrounding the shell is part of the Sgr A East halo.}
\end{figure}

\begin{figure}
\caption{(left) A VLA radio continuum image of the inner 20$''$ of the
Galactic center at $\lambda$=1.2cm with a resolution of
$0.5''\times0.3''$ is displayed in blue and is superimposed on stellar
distribution at $\lambda$=2 micron (96) displayed in red with a
cellsize of 50 milliarcseconds.  (right) A schematic digaram showing
major stellar and gaseous components of the Galactic center.}
\end{figure}

\begin{figure}
\caption{The bow shock structure and the ionized tail of a cometary
feature associated with IRS 7 at $\lambda$=1.2cm (blue) with a
resolution of 0.3$''\times0.2''$ is superimposed on stellar emission
from the photosphere of IRS 7 at 2$\mu$m at the apex of the bow shock
(yellow spot) based on the VLA and NTT observations, respectively.  A
precise alignment of these images has been made possible because
a number of SiO and H$_2$O masers in radio wavelengths have been
identified with stellar sources in the diffraction limited 2.2 $\mu$m
infrared images to within 30 milliarcseconds (96).}
\end{figure}

\begin{figure}
\caption{Schematic diagram showing the relative positions and
sizes of the Halo and Sgr A East relative to Sgr A*, which is
shown here as a point centered within the CND. The thermal three-arm
spiral radio source Sgr A West is also contained within the CND.
Low-energy photons arising from the CND are also shown to be upscattered to
higher energies
by relativisitic particles of Sgr A East. }
\end{figure}

\begin{figure}
\caption{The broadband spectrum calculated self-consistently using the 
particle
decay products from proton-proton scattering. The electron and positron
distributions
result from the decay of pions produced in these scattering events.  The data
included
in this plot are from reference (38, 89, 97, 98).}
\end{figure}

\begin{figure}
\caption{The brightest region of continuum emission from Sgr A East at
$\lambda$=20cm in blue with the bright white spot coincident with Sgr
A$^*$ is superimposed on the brightest portion of diffuse X-ray
emission from the Galactic center presented in red (89).  The bright
red spot to the southwest corner is associated with a low-mass
eclipsing X-ray binary system (99).  The crosses represent position
of OH masers
at 1720 MHz signifying the region where a molecular cloud is
interacting with the
nonthermal shell of Sgr A East (36).  }
\end{figure}

\begin{figure}
\caption{Contours of radio continuum emission at 6cm (green) as seen in
Figure 4 are
superimposed on the distribution of H$_2$ emission using  
the Hubble Space Telescope near-infrared camera and multiobject spectrometer
(NICMOS).  The
white crosses indicate OH (1720 MHz) maser positions, whereas the yellow
cross coincides with Sgr A$^*$.
The two crosses toward the top left and right are maser B and C lying on the 
CND and the  H$_2$2 filament at 134 and 43 \kms. Contours of Sgr A West
lie within the brightest  H$_2$ emitting features delineating the CND.
(39). see also an additional figure on
WWW.sciencemag.org/feature/data/1046905.shl or
http://www.physics.nwu.edu/research/zadeh.html under Science Online Article}
\end{figure}

\end{document}